\documentclass[a4paper,11pt]{article}

\usepackage{siunitx}

\usepackage{float}
\usepackage{ctable}
\usepackage{subcaption}
\usepackage{jinstpub} 

\usepackage{lineno}

\title{\boldmath First 3D reconstruction of a blast furnace using muography}

\author[a,b,c]{A. Cohu}
\author[c]{A. Chevalier}
\author[b]{O. Nechyporuk}
\author[d]{A. Franzen}
\author[d]{J. Sauerwald}
\author[a,c]{J-C. Ianigro}
\author[a,c]{and J. Marteau }

\affiliation[a]{IP2I, IN2P3, CNRS, Université Lyon 1, UMR {5822}, \mbox{69100 Villeurbanne, France}}
\affiliation[b]{ArcelorMittal Maizières Research SA,Voie Romaine, 57280 Maizières-lès-Metz, France}
\affiliation[c]{{MUODIM,} 
 31 rue Saint-Maximin, 69003 Lyon, {France}}
 \affiliation[d]{ArcelorMittal Bremen GmbH, Carl-Benz-Straße 30, 28237 Bremen, Germany}

\emailAdd{marteau@ip2i.in2p3.fr}

\abstract{The blast furnace (BF) is the fundamental tool used in the iron manufacture. Due to the difficulty of making direct measurements of the inner state of blast furnaces, we determined the density distribution of its internal volume in order to improve its productivity using muography. This is an imaging technique based on the differential absorption of a flux of incident particles, muons, by the target under study, similar to clinical X-ray imaging. Muons are elementary particles that have the property of passing through dense materials, up to hundreds of meters of rocks. Their relative absorption and deviation allows the generation of density distribution images of an object by tracking the number of muons received by a detector, before and after passing through a structure. The incident direction of the detected muons is reconstructed by means of a detector composed of 3 scintillator panels that we moved on 3 positions around the BF. With this technique, we obtained the first 3D image of the internal structure of a BF using a Markov Chain Monte Carlo (MCMC) inverse problem solving algorithm on muon flux data. We were also able to perform density monitoring of the BF and monitor some of its operating parameters. We distinguished the position and shape of the cohesive zone, a key element in the productivity of a furnace, validating this innovative measurement concept in the application to a BF and opening the field to a series of future experiments to gain both spatial and temporal resolution. }


\arxivnumber{} 

\begin{document}
\maketitle
\flushbottom

\section{Introduction} \label{sec:intro}
We seek to determine the density distribution of matter inside a blast furnace in order to visualize the cohesive zone and to carry out a dynamic monitoring of the various phases present in the blast furnace. To achieve this, muography is applied to make a dynamic image during the operation of a blast furnace of ArcelorMittal in Bremen, Germany. Muography measures the absorption or deflection of cosmic muons as they pass through dense materials. Muons are elementary particles which have the property to pass, in a straight line to first order, up to several kilometers of standard rocks, and whose relative absorption allows to generate images by contrast densitometry, like a standard clinical radiography. The acquired muon data allows to follow the density as a function of time during the operating cycles of the blast furnace. In a second step, the acquisition of 2D images and then the 3D reconstruction is accomplished by the data inversion from several measurement points. The final objective is to understand the topological characteristics and the formation rate of the cohesive zone and the influence of certain loading parameters (blast pressure, coke rate at the cohesive zone ...). 

Muography experiments have already been performed on BF \citep{takamatsu,vanini}. Nagamine et al.\citep{nagamine} measured the thickness of the brick used for the base plate and side wall in 45 days. The aim was to predict the lifetime of the furnace which is usually 20 years with a consumption of bricks a few meters thick, equivalent to about 1 cm/month. In addition, the local density (at ± 0.2 g/cm 3) and the time-dependent behavior of the iron-rich part were determined. They observed a change in the internal structure of BF after a 1.5 day stop of the hot air supply. Hu et al.\citep{hu} used muography to distinguish iron pellets from coke. They found that the measured linear scattering densities (LSDs) correlated linearly with the measured material densities. However, the results presented in this article are the first reconstructions in 3 dimensions, has never been done. There are papers presenting 3D structures of volcanoes \citep{rosas,Nagahara} obtained with muon tomography with similar inversion challenges but it's the first time on a BF.

The article is organized in four main parts. The first part explains the theoretical points and algorithms used in tomography reconstructions. The second part presents the simulation parameters of a blast furnace and explains how we built the 3D image. We have completed our analysis by monitoring the activity of the blast furnace as function of different environmental parameters such as atmospheric pressure and temperature. The third part presents the results of real muons data inversion to visualize the different density zones in the blast furnace. Finally, we report the conclusions and perspectives of this study.

\section{Tomography reconstruction theory} \label{sec:2}
\subsection{Generalities and issues}
Absorption muography measures the cosmic ray flux deficit in the direction of observation and determines the integrated density of a structure. Muons are cosmic rays able to cross several meters of rocks losing energy. The minimum amount of energy that muons need to penetrate the structure must be of a value higher than the one lost inside the object, so that the detector can follow the outgoing muons. Position must then be adjusted to optimize the spatial resolution. Muon tomography is limited to the study of a portion of the object only, because of the limited angular aperture of the detector. In order to produce a 3D image, the detector need to be placed in different locations. 

The \textbf{direct problem} consists, in our case, of predicting the expected muon flux at the exit of an object. It is necessary to use materials knowledge and physical properties of the object's constituents and then to use a density distribution as precise as possible. The parameters sought during the inversion use the direct problem to estimate the expected measurements. The measured information is retrieved from a given distribution of $p$ parameters in the studied structure. Moreover, the flux attenuation is estimated from a known law based on the contrasting distribution of zones (of different density for example).

With the trackers the direction of the muons is reconstructed (see subsection \ref{section3pt2}) in order to observe the properties of the medium on a precise observation axis (see Lesparre et al.\citep{lesparre_design_2012}). The flux that arrives on the detector after having crossed the object and the theoretical flux that would reach the detector in the absence of matter are compared. The contrast between thoses two quantities gives directly access to the \textbf{opacity} of the matter (in meter equivalent water (mwe)), defined as the integral of the density along the trajectory of the muon from its entry point to its exit point. Moreover, the only observables are the particles directions and energy deposits, since the detectors usually do not give direct information on the particles total energy. In order to solve the data inversion, the measured flux is coupled to a theoretical flux model and to a flux loss model in matter.

 \subsection{Inverse problem \label{inv_probl}}
 
The reconstruction of a muography image is achieved by solving an inverse problem \citep{tarantola2005inverse}. The goal is to recover the distribution of properties of the medium (\textbf{3D density}) from measurements of muon rates and their directions. An inverse problem is a situation in which one tries to determine the parameters of a model $p$ (here the 3D density of the environment) from experimental measurements $m$ (muon rates) such that $m=f(p)$ where $f$ contains the open muon flux (calibration) and the law governing the absorption of muons in matter. While a single parameter estimate can be easily obtained by least squares fitting, the use of Monte Carlo methods allows the maintenance of the stochasticity of the model during the estimation of these parameters. In order to improve the reliability of the results, it is good practice to add some information about the object to study, other than data, that we call \textit{a priori} information. This allows to constrain the solutions space and improve the accuracy of the statistical answer.



We use the Metropolis-Hastings algorithm which is a particular class of Monte Carlo methods using Markov chains (MCMC). It works like a geometrically biased random walk with a data based selection at each throw \citep{sambridge2002monte}. Each new proposal/model can be accepted or rejected if the likelihood of the model (regarding the data and the physics of the problem) is greater or smaller than the likelihood of the previous model. Hence, unlike more general Monte Carlo methods where the sampled values are statistically independent, in the Metropolis-Hastings algorithm they are statistically auto-correlated. This auto-correlation is minimized by adding a threshold number of changes that must be accepted between the recording of two consecutive samples. In practice, we should try to record different models. We voxalized the volume and modified its content by changing 400 voxel values per attempt, with statistical recording every 10 models. That corresponds to 4000 changes of voxels between two models. It is an empirical criterion defined manually to decorrelate the semblance of two models, while having statistics in a reasonable time.

\begin{figure}[H]
   \centering
  \includegraphics[width=\textwidth,trim=0cm 0cm 0cm 0cm,clip=true]{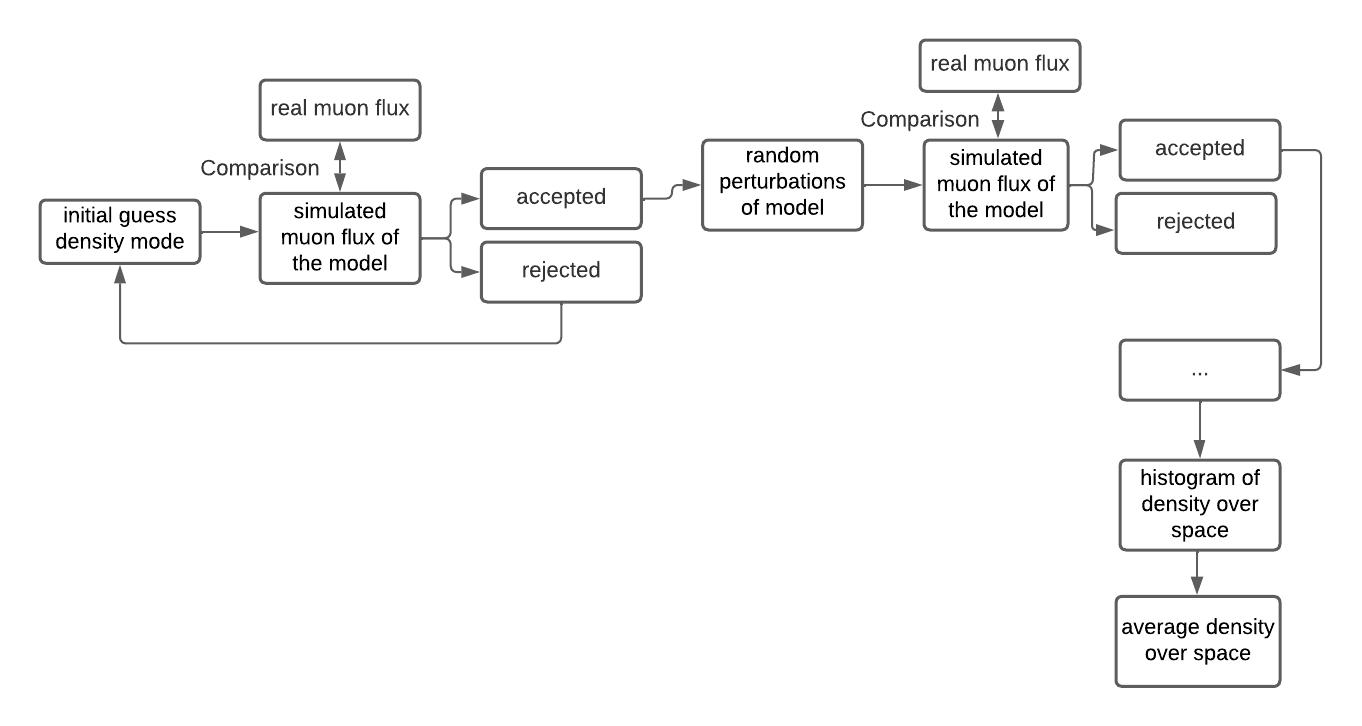}
	\caption{MCMC model algorithm workflow. }
	\label{exlication_MCMC}
	
\end{figure}

 We need models whose density values are continuous over finite element discretized volumes. Here a model stands for any given set of values representing a physical system. The engine that generates the 3D density models, linked by Markov chain, is made to design density sets per class forming spatially contiguous voxel sets which share the same density. The inversion algorithm is a 3D adaptation of Mosegaard et al.\citep{mosegaard1995monte} work, where details can be found in Chevalier et al.\citep{chevalier2014monte}. A workflow of our MCMC model algorithm based on their work is presented in Figure \ref{exlication_MCMC}. The method makes selections of the models according to the evolution of the distance $D$ :
\begin{equation}
 D= \frac{F_t - F_m}{\sigma_m} 
 \label{D}
\end{equation}
with $F_t$ the theoretical flux, that depends on our assumed material content for each voxel, and $F_m$ the measured flux. $D$ is a metric of the distance between the data and the simulation, expanded or compressed by the degree of uncertainty on the measurement $\sigma_m$. This distance $D$ is recalculated each time the density changes. A new model is accepted as a solution if the mean difference between the reconstructed signal and the data (evaluated by the mean square deviation) is lower than the mean noise estimated during the measurement. If the model is selected, we save its likelihood. This model is then slightly perturbed again by changing some voxel density values and we recalculate the new likelihood. The total likelihood of a model, $L(m)$, can be expressed as a product of partial likelihoods, one for each data type. 
 
 Our inversion algorithm is able to couple information from several detectors at the same time. Indeed, each detector measures a flux and \textit{travel length} (defined as the thickness of material seen by the detector in its acquisition configuration) per viewing axis. By taking a set of density distributed in the voxels, we have the opacity along the line of sight. The number of registered models can be modified and it depends on the requirements in terms of accuracy and available computing time. 

\section{Application to a blast furnace}

 \subsection{Operating constraints} 
\begin{figure}[H]
   \centering
  \includegraphics[scale=0.5,trim=0cm 5cm 0cm 5cm,clip=true]{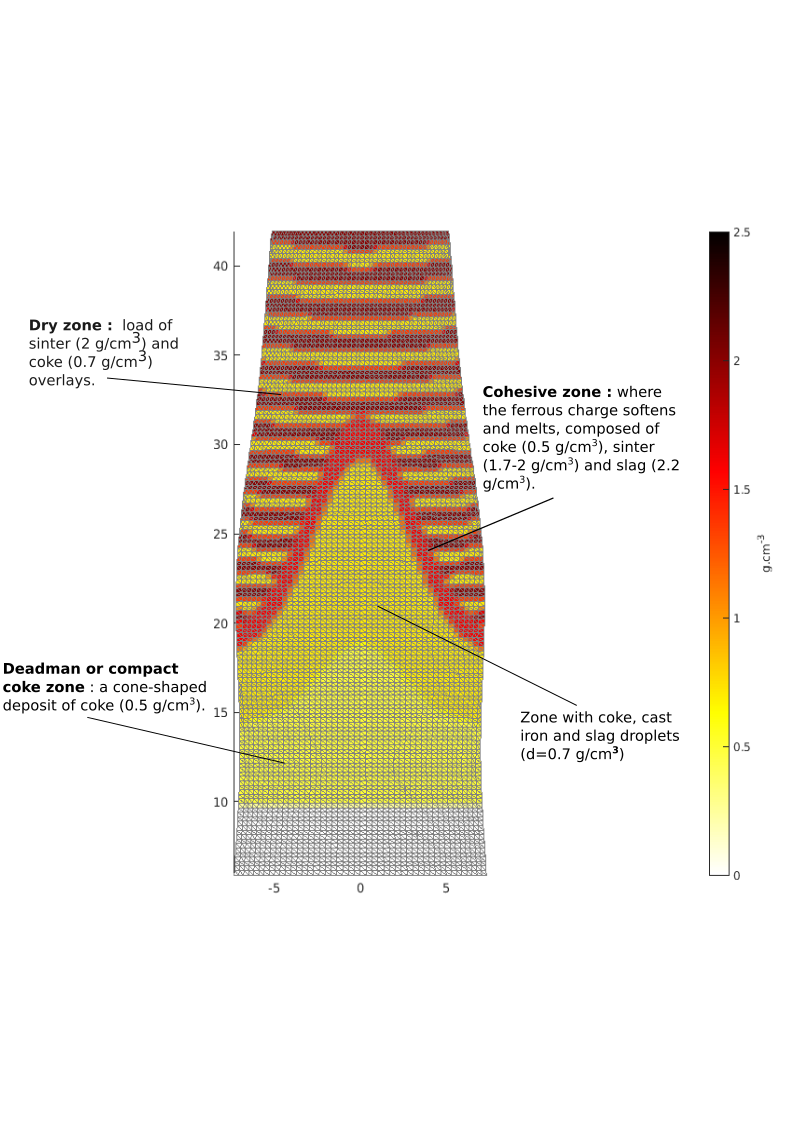}
	\caption{Density pattern in the different blast furnace zones, distributed in its meshes (unit : meter).}
	\label{drawing_hf_zone}
	
\end{figure}

The blast furnace we worked on is a structure of about 50 m high and 15 m wide. Its internal components include coke and cast iron with different densities (0.5 \si{t/m^3} and 6.8 \si{t/m^3} respectively). Its operating process is complex and involves several chemical and physical reactions. A sketch of a BF is presented in figure \ref{drawing_hf_zone} with differents functions and densities zones (dry zone, deadman, cohesive zone ...). We used these previous details as a priori information for our analysis.

 \subsection{Acquisition configurations \label{section3pt2}}

\begin{figure}[ht]
	
	\centering
 \includegraphics[scale=0.18,trim=9cm 0cm 10cm 0cm,clip=true]{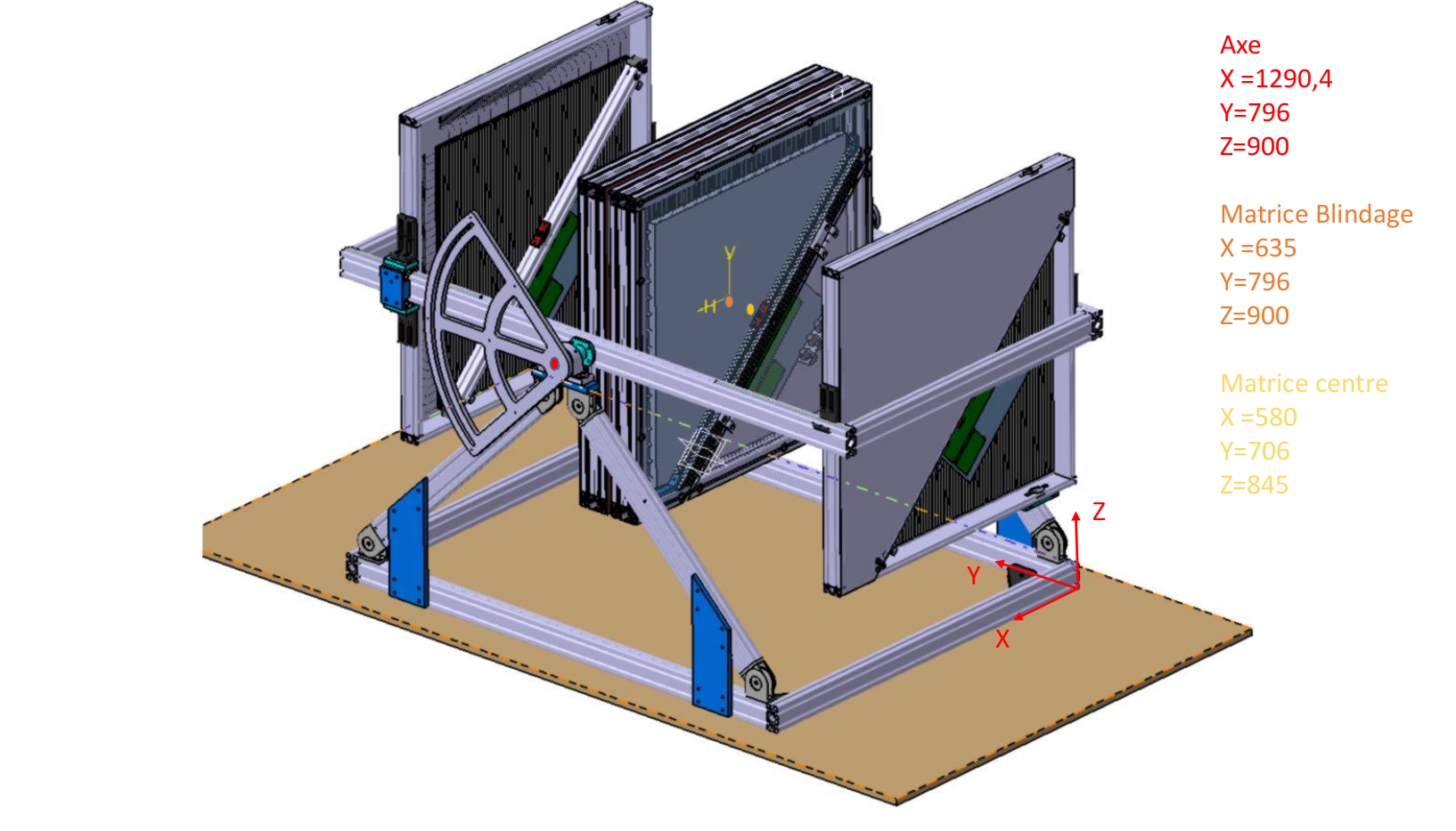}
	\includegraphics[scale=0.4,trim=10cm 5cm 9cm 3cm, clip=true]{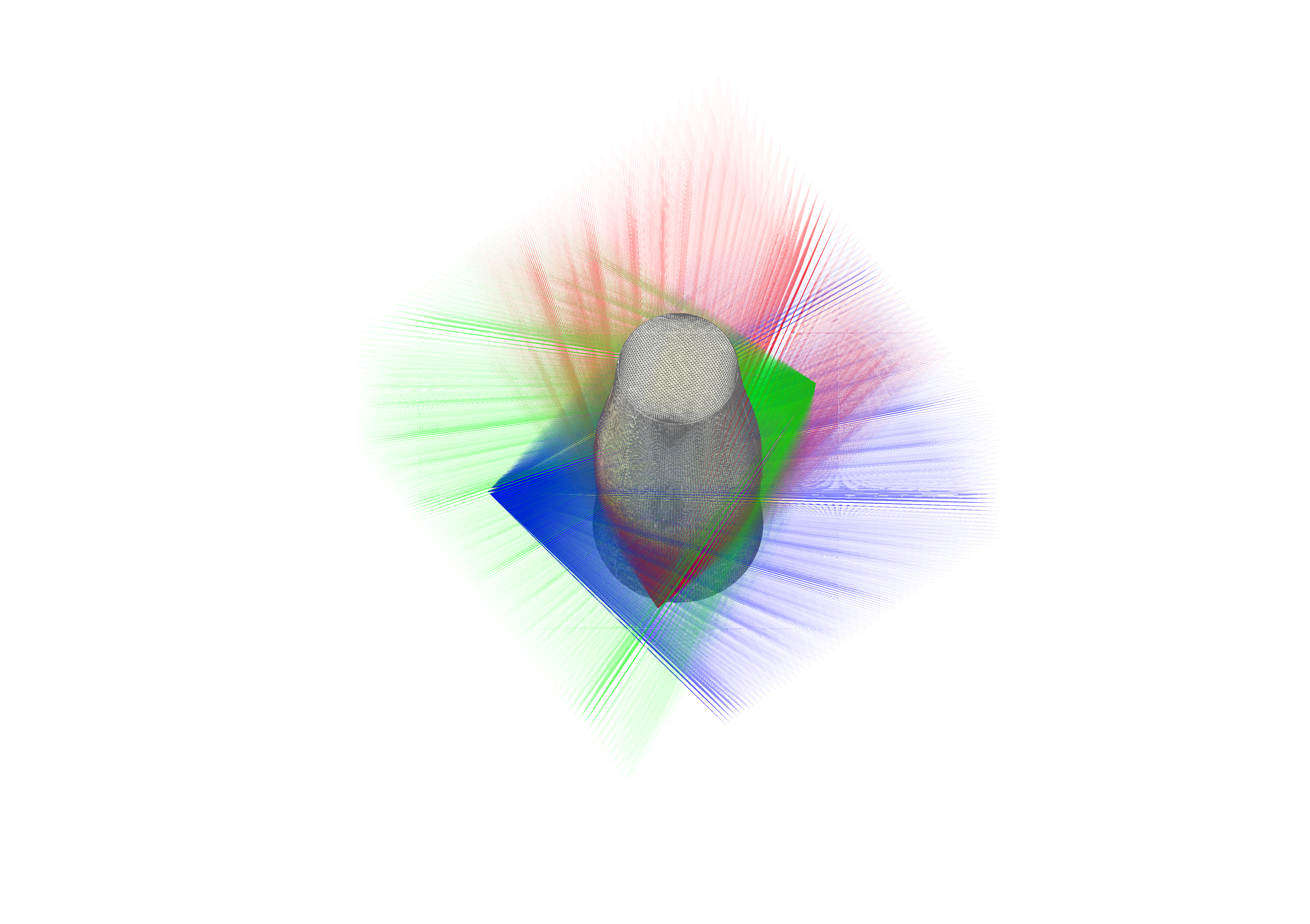} 
	\caption{\textit{On the left:} A sketch of the detector system with scintillator planes of 80 by 80 \si{cm^2} and a distance between the first and the last plane equals to 120 cm. \textit{On the right:} Detector virtual lines of sight at each position with different colors : position/run 1 (\textit{in red}), position 2 (\textit{in blue}), position 3 (\textit{in green}). }
	\label{emplacements_detecteurs_TL}
\end{figure}

Muons data presented in this article were taken with a scintillators 3-planes muon tracker built in Lyon, IP2I laboratory, with an angular resolution of 1.2$^\circ$ (see the sketch in figure \ref{emplacements_detecteurs_TL} left). This detector has already been used in previous muography experiments: in Greece \citep{avgitas}, Italy \citep{acernese2022virgo} and Guadeloupe \citep{marteau_muons_2012,jourde}. Multi-tracks are suppressed, electrons are partly suppressed by the multiplicity cuts on the detection planes, but there is no PID in the system. 

  In order to obtain a 3D image of the blast furnace, we carried out three muography acquisitions around the blast furnace, performed between the end of July 2021 and the end of March 2022 with the 3-planes detector (specifications of the runs are described in table \ref{charact}). In this way we observe a maximum volume of the BF (and not only an area), common to the fields of view of the three detectors. In figure \ref{emplacements_detecteurs_TL} right, the detector virtual lines of sight at each position are shown. They allow visualization of the field of view of each location and the common areas that are observed.

\begin{table}[ht]	
\centering
\caption{Specifications of the 3 runs shown in figure \ref{emplacements_detecteurs_TL} (Ze=zenith, Az=azimuth angles). \\ 
}

\begin{tabular}{|c|c|c|c|c|c|c|}
\hline
\textbf{Run} & \textbf{Ze (°)} & \textbf{Az (°)} & \textbf{X (m)} & \textbf{Y (m)} & \textbf{Z (m)} & \textbf{Dates}                                                                      \\ \hline
\textbf{1}    & 45              & 51              & -7.51          & -7.04          & 11.35          & 07/29/21 - 20/09/21                                                                 \\ \hline
\textbf{2}    & 60              & 144.33          & -13.49         & 9.52           & 14.55          & \begin{tabular}[c]{@{}c@{}}10/08/21- 10/21/21\\ \& 12/21/21 - 01/24/22\end{tabular} \\ \hline
\textbf{3}    & 60              & 270             & 15.14          & 0.24           & 14.55          & 02/03/22 - 03/31/22                                                                 \\ \hline
\end{tabular}

		\label{charact}
\end{table}

\subsection{External and internal parameters effects on muon flux}

When counting the muons rate, in a given direction as a function of time for each of the positions shown in figure \ref{emplacements_detecteurs_TL}, we observe temporal variations. The rate of high-energy cosmic ray muons as measured underground is shown to be strongly correlated with upper-air temperatures during short-term atmospheric phenomena  \citep{adamson2014observation,tramontini2019middle} and with pressure \citep{jourde_monitoring_2016}. We have evaluated the effects of atmospheric parameters and those of the target operation (coke rate estimate at the cohesive zone and the value of the blast pressure, defined above) by performing a linear fit and calculating the correlation coefficient between muons rate and parameters. Our goal is to substract these different effects from the measured rate, before realizing the data inversion, to obtain a density distribution 3D image in the blast furnace.

\begin{figure}[H]

	\centering
	\includegraphics[width=0.92\textwidth]{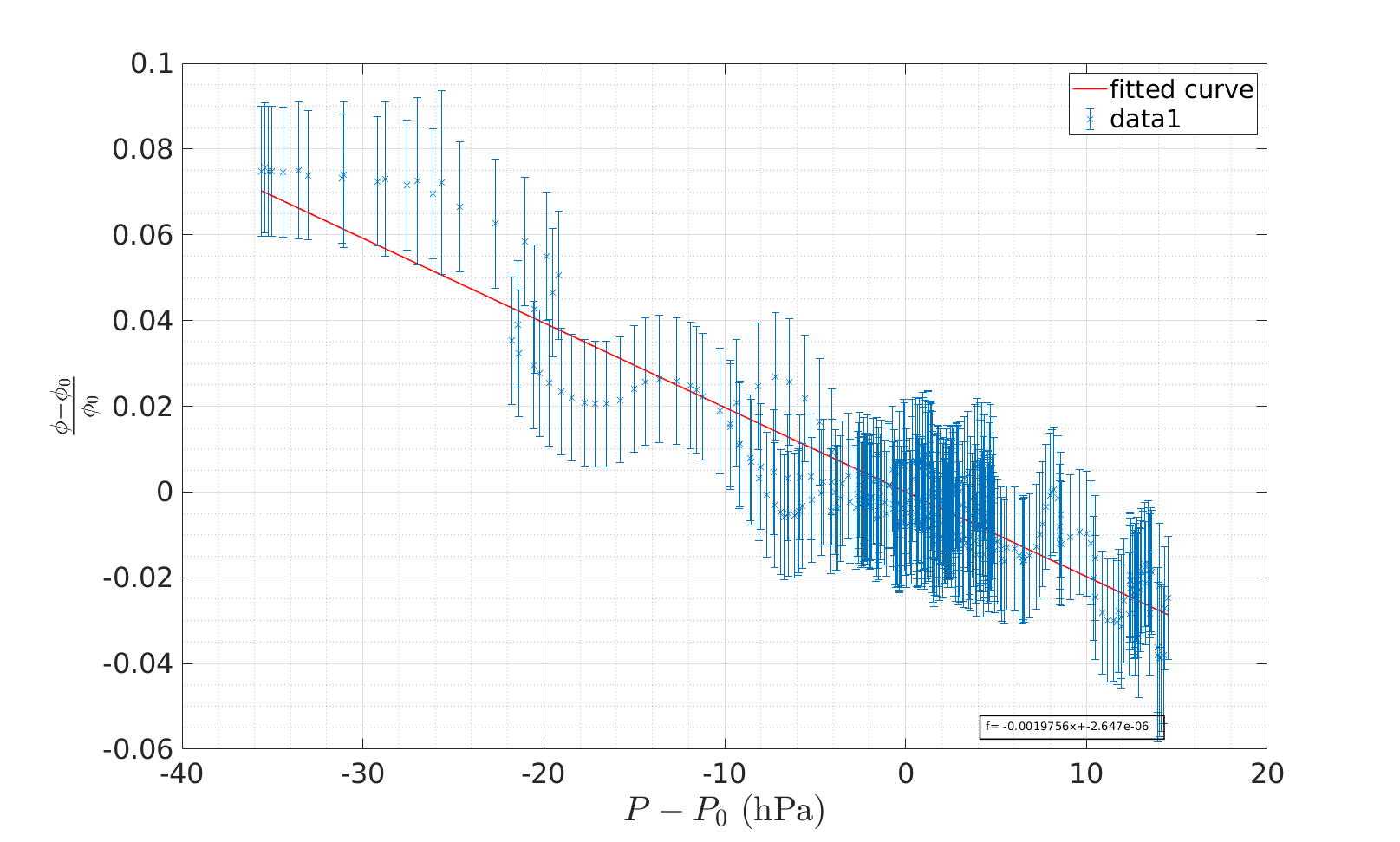}
	\caption{Relative muon flux as a function of the deviation from the mean of the pressure $P-P_0$. The errors associated with the fluxes are standard deviations fixed at 2$\sigma$. }
	\label{fit_ambiant_pressure}
	
\end{figure}

We chose to compute the pressure/flux correlations for each position separately because the detectors don't see the same BF opacity/zone from one location to another. Depending on the direction, the opacity is different. By performing a linear fit between the flux and ambient pressure data (see an exemple in figure \ref{fit_ambiant_pressure}), we obtain the barometric coefficients $\beta_p$ (\si{hPa^{-1}}), such as \begin{equation}
    \dfrac{(\phi-\phi_0)}{\phi_0}= \beta_p \times (P-P_0)
\end{equation} 
(see Jourde et al.\citep{jourde_monitoring_2016}) where the left-hand side represents the relative muon flux (with $\phi_0$ the average flux), $P$ the pressure, $P_0$ the average pressure over the duration of the experiment and $\beta_p$ a linear coefficient in \si{hPa^{-1}}. Their error (uncertainty) as well as the correlation coefficient of the fit are calculating using a Matlab function and are presenting in the table \ref{tomo_coefP}. The correlation values between pressure and flux vary with time according to the table \ref{tomo_coefP} and the sensitivity of the muon flux to the pressure variation appears more important in high pressure episodes. Moreover, the muon fluxes do not seem to be sensitive to the temperature variations of the upper atmosphere over the studied periods and they are thus more affected by the pressure variations. This result is consistent with the opacity values we encountered, at most 50 mwe (lower than those of Tramontini et al.\citep{tramontini2019middle} $\sim$ 700 mwe and close to Acernese \citep{acernese2022virgo} open sky experiment) which do not allow to filter out the low energy muons. Only the most energetic muons are sensitive to temperature variations.\\

\begin{table}[ht]
\centering
\caption{Barometric coefficient, their errors/uncertainties and the correlation coefficients of the different acquisitions around the blast furnace.\\}
\begin{tabular}{|c|c|c|c|}
\hline
\textbf{Run} & \textbf{$\beta_p$ (\si{hPa^{-1}})} & \textbf{Error (RMS)} & \textbf{Correlation coefficient} \\ \hline
\textbf{1}        & -0.0011                & 1.3$\%$                & 0.77                             \\ \hline
\textbf{2}        & - 0.0015               & 0.9$\%$                & 0.80                             \\ \hline
\textbf{3}        & - 0.0018               & 1$\% $                 & 0.47                             \\ \hline
\end{tabular}

\label{tomo_coefP}
\end{table}

 A blast furnace is considered to be in operation when air is injected into it, measured by the value of the so-called blast pressure. At standstill, the density in the blast furnace is greater and the interior "column" is tighter. In addition, as the blast pressure increases, fine particles of sinter take the place of gas and the density in the BF increases. Furthermore, a high coke fraction in the cohesive zone means that the associated density is lower than usual. Several parameters can therefore affect the muon flux by changing the density inside: the fraction of fine particles, the blast furnace stop and the addition of coke.  

We performed multivariate linear fits (with external : pressure $P$ and temperature $T$ and internal parameters : coke rate estimate at the cohesive zone $CR$ and blast pressure $BP$) on the relative muon flux and evaluated the adequacy of our fits with the Pearson linear coefficient of determination. Results are presented in table \ref{coef_pearson}. The pressure appears to be the dominant parameter and Pearson coefficients don't change a lot by adding more variables. 
\begin{table}[ht]
\centering
\caption{The "3 acquisitions" Pearson coefficients with an increasing number of variables/parameters present in the linear regression performing on the relative muon flux. P= Pressure, T=Temperature, CR= coke rate and BP= Blast Pressure.\\}
\begin{tabular}{|c|c|c|c|}
\hline
\textbf{Run} & \textbf{P + T} & \textbf{ P + T + CR} & \textbf{P + T + CR + BP} \\ \hline
\textbf{1}         & 0.45                          & 0.45                                                 & 0.571                              \\ \hline
\textbf{2 (Oct)}       & 0.831                                                                                                               & 0.861                                                                                                                        & 0.867                            \\ \hline
\textbf{3}              & 0.786                                                                                                               & 0.788                                                                                                                        & 0.791                         \\ \hline
\end{tabular}

\label{coef_pearson}
\end{table}

\begin{figure}[H]

	\centering
	\includegraphics[width=\textwidth]{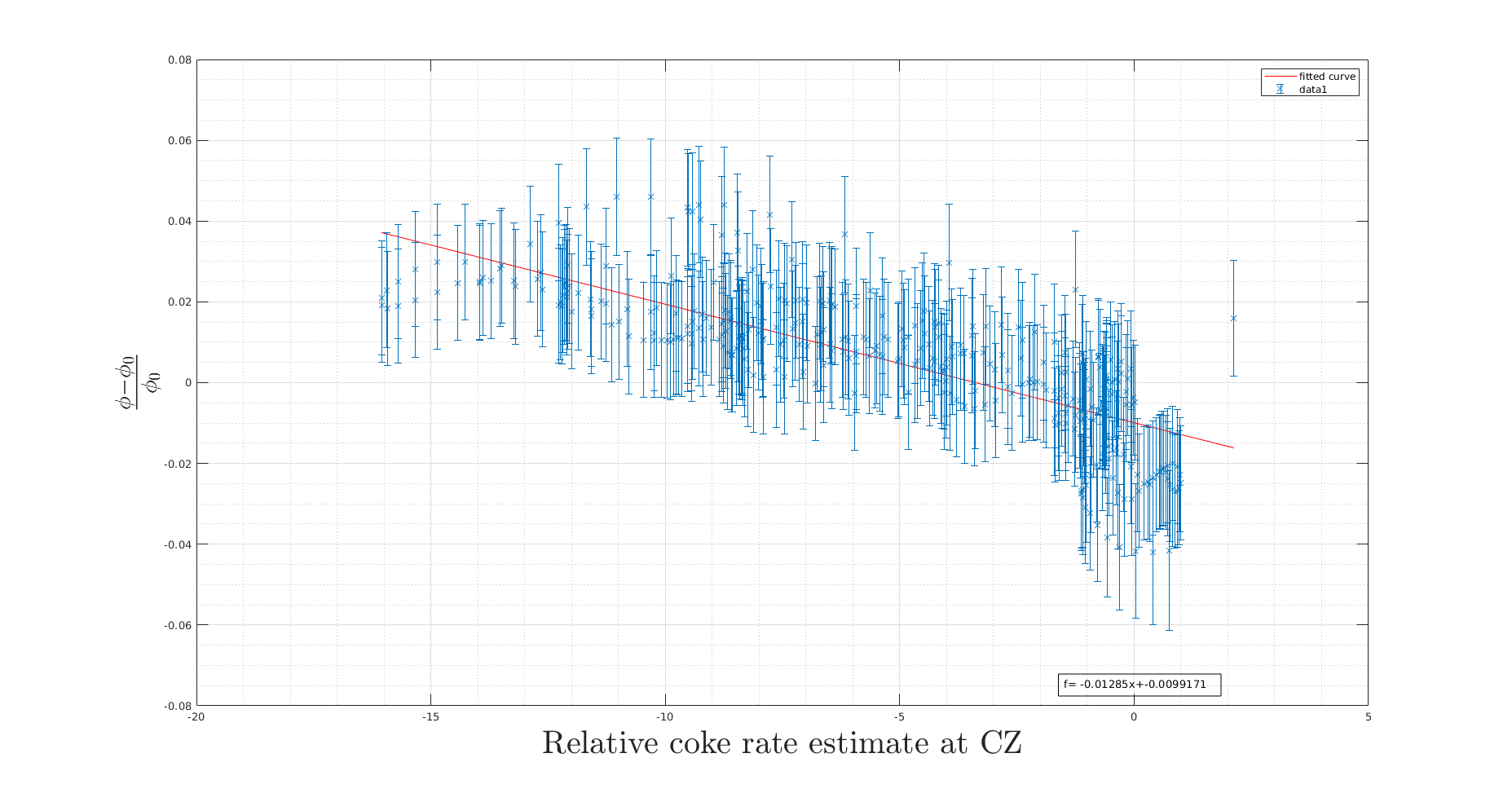}
	\caption{Adjustment of the relative muon flux that passed through the blast furnace and corrected for pressure, with the relative estimated coke rate in the cohesive zone for Run 2.}
	\label{cokeratefit}
	
\end{figure}

Moreover, in October 2021, during a high pressure period ( $\geq$1020 mbar), the coke rate in the cohesive zone seems to be well correlated with the pressure corrected muon flux (see figure \ref{cokeratefit}). We found $\gamma_{CR}$=-0.013 ($\pm 1\%$) with a correlation coefficient of 0.76. This means that when the coke rate is high, the blast furnace is stopped and the material goes down, as well as the cohesive zone, so the density in the blast furnace increases and the measured muon flux behind it decreases.

  After solving the direct problem we reconstruct the 3D-average-density model of the blast furnace by using the measured flux (corrected from parameters that affect it). The Markov chain Monte Carlo method described in subsection \ref{inv_probl} is used.

\section{Results} \label{sec:3}
The results presented in this section were obtained using real data measured by the detector for three runs.

\subsection{2D fields}

\begin{figure}[]
     \centering
      \begin{subfigure}[b]{0.45\textwidth}
         \centering
         \includegraphics[scale=0.45,trim=4.7cm 1.7cm 0cm 0.5cm,clip=true]{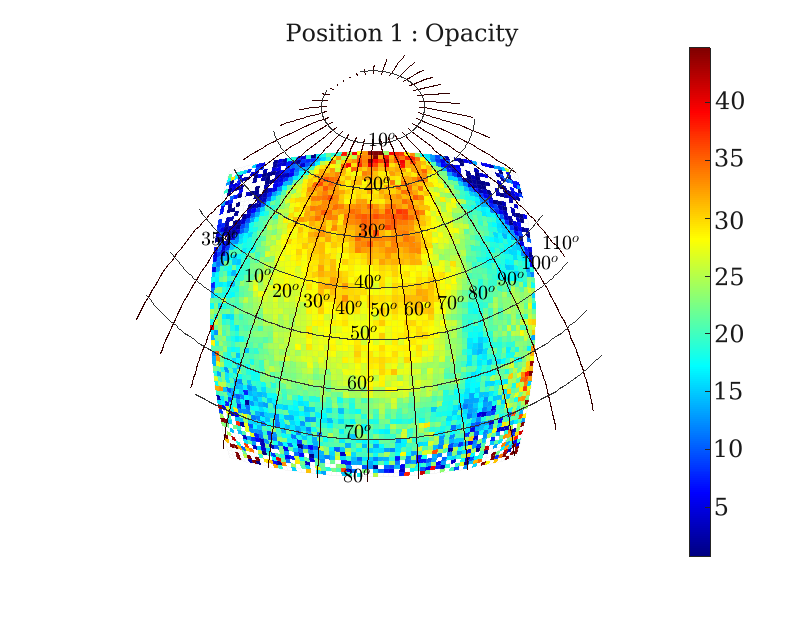}  
         \label{fig:three sin x}
     \end{subfigure} 
         \begin{subfigure}[b]{0.45\textwidth}
         \centering
         \includegraphics[scale=0.45,trim=3cm 1.7cm 1.2cm 0.5cm,clip=true]{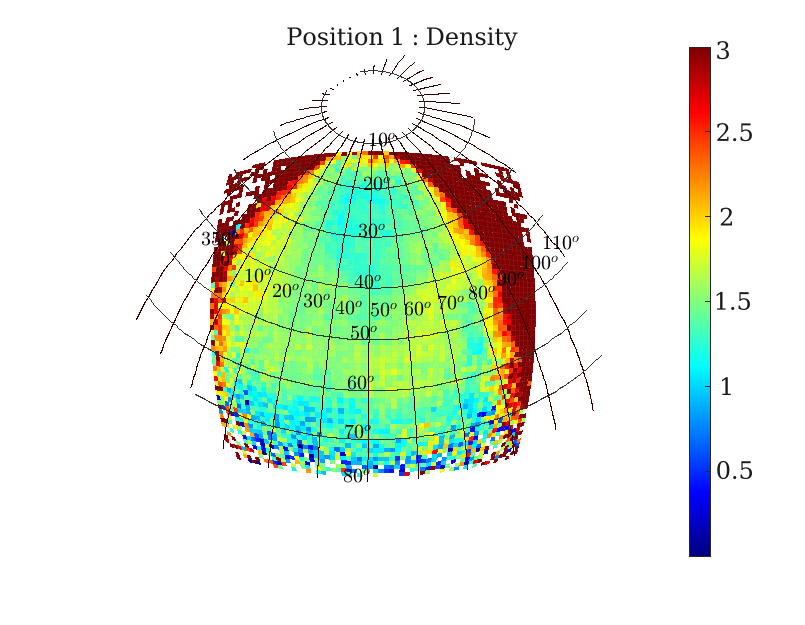}  
         \label{fig:y equals x}
     \end{subfigure}
     \hfill

     \begin{subfigure}[b]{0.45\textwidth}
         \centering
         \includegraphics[scale=0.45,trim=4.7cm 1.7cm 0cm 1cm,clip=true]{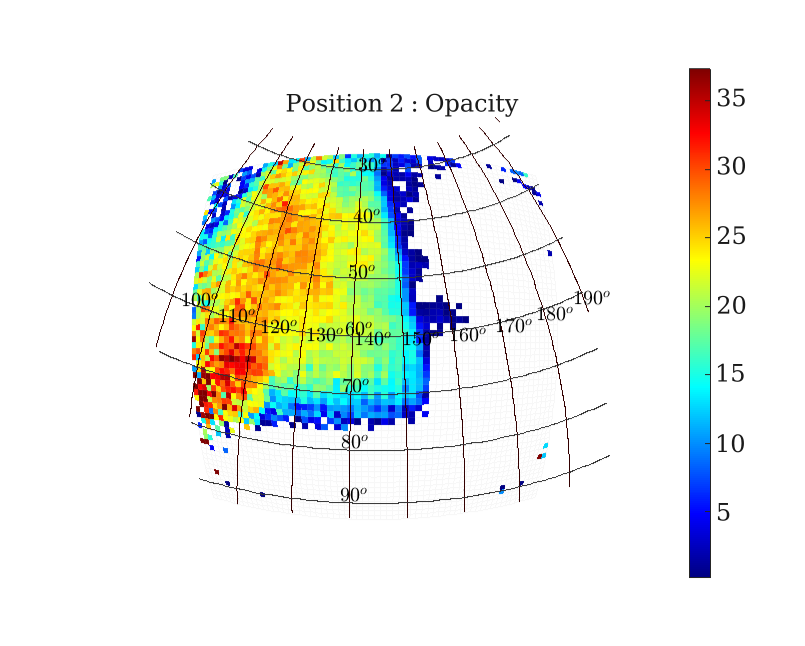}  
         \label{fig:three sin x2}
     \end{subfigure}  
                \begin{subfigure}[b]{0.45\textwidth}
         \centering
         \includegraphics[scale=0.45,trim=3cm 1.7cm 1.2cm 1cm,clip=true]{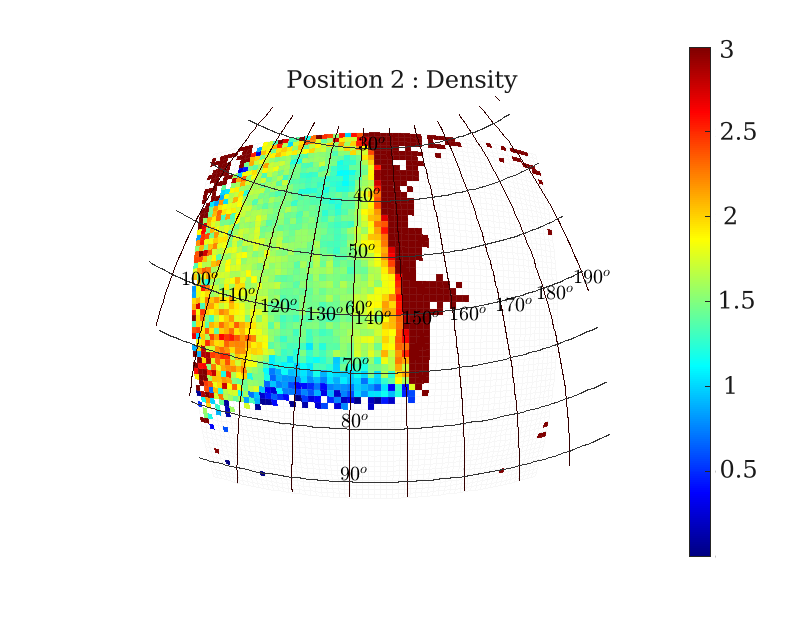}  
         \label{fig:y equals x2}
     \end{subfigure}
   
     \hfill
     \begin{subfigure}[b]{0.45\textwidth}
         \centering
         \includegraphics[scale=0.44,trim=4.4cm 2.2cm 1.5cm 0cm,clip=true]{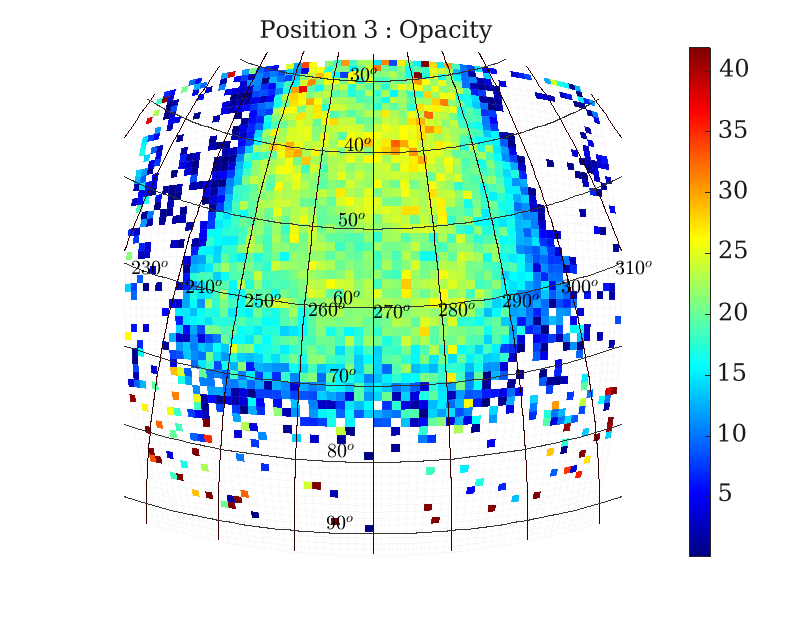}    
         \label{fig:three sin x3}
     \end{subfigure}
     \hfill
     \begin{subfigure}[b]{0.45\textwidth}
         \centering
         \includegraphics[scale=0.14,trim=13.5cm 7cm 0cm 1.5cm,clip=true]{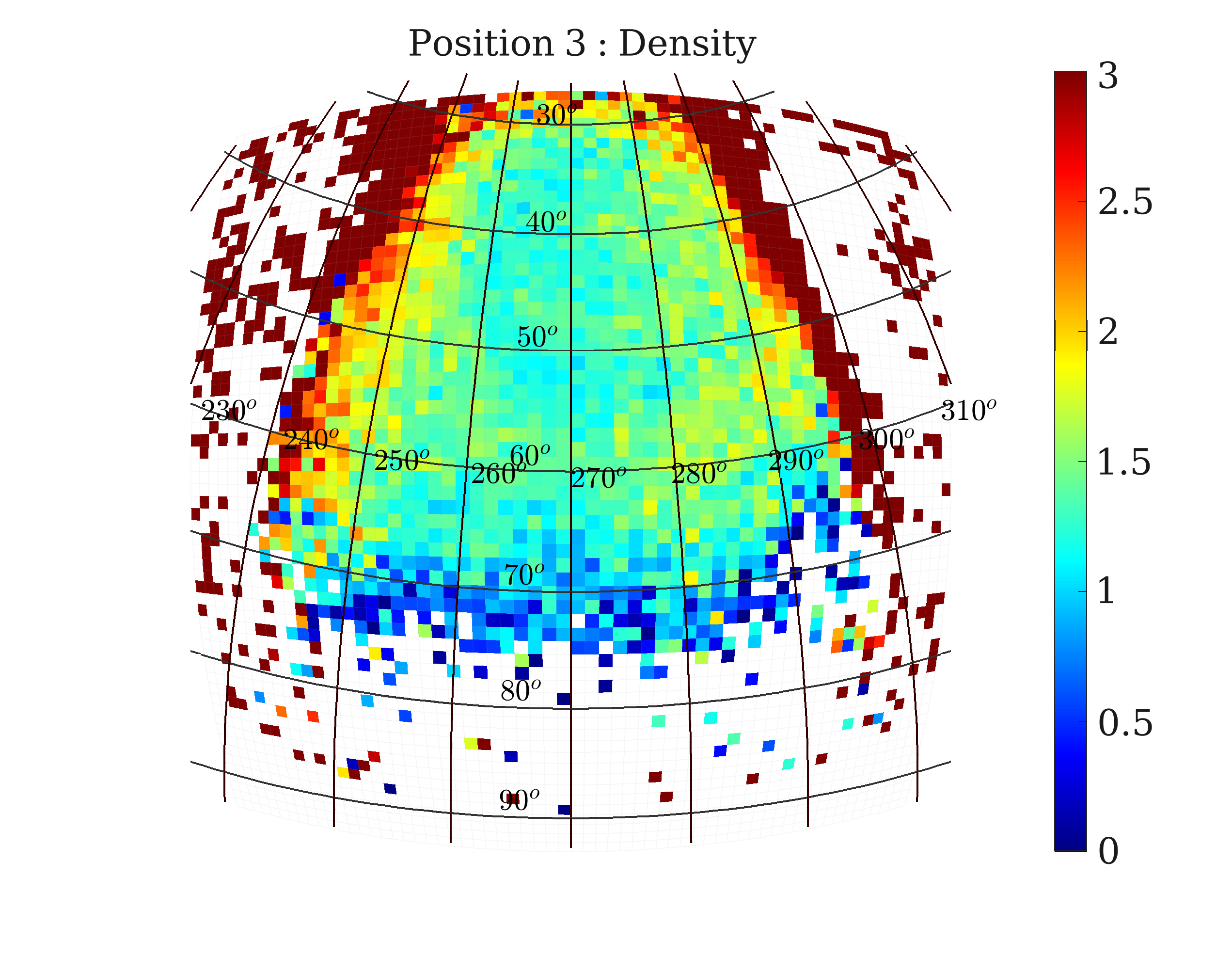}   
         \label{fig:y equals x3}
     \end{subfigure}

         \caption{\textit{On the left}, the \textbf{opacity} (in mwe) and \textit{on the right}, the \textbf{apparent density} (in \si{g.cm^{-3}}) for each of the 3 positions are represented. The axes are the azimuth and zenith in $^\circ$.}
         \label{opacity_density}
\end{figure}

 In the figure \ref{opacity_density} the opacities (left panels) and densities (right panels) seen by the detector at each of its 3 positions are represented in 2D, before inversion.  We can see a slightly denser area (in yellow) in the middle of the figures. This zone would seem to be a 2D projection of the cohesive zone. The shells of the blast furnace are clearly visible on density representations. The position 2 density figure shows a denser area in the center left.  As expected, no muons are recorded below 75-90° in the data. Finally, the information contained in these figures is aggregated and inverted to reconstruct the 3D blast furnace and the density distribution inside.

\subsection{3D-inversion results}
\begin{figure}[]

	\centering
	\includegraphics[width=\textwidth]{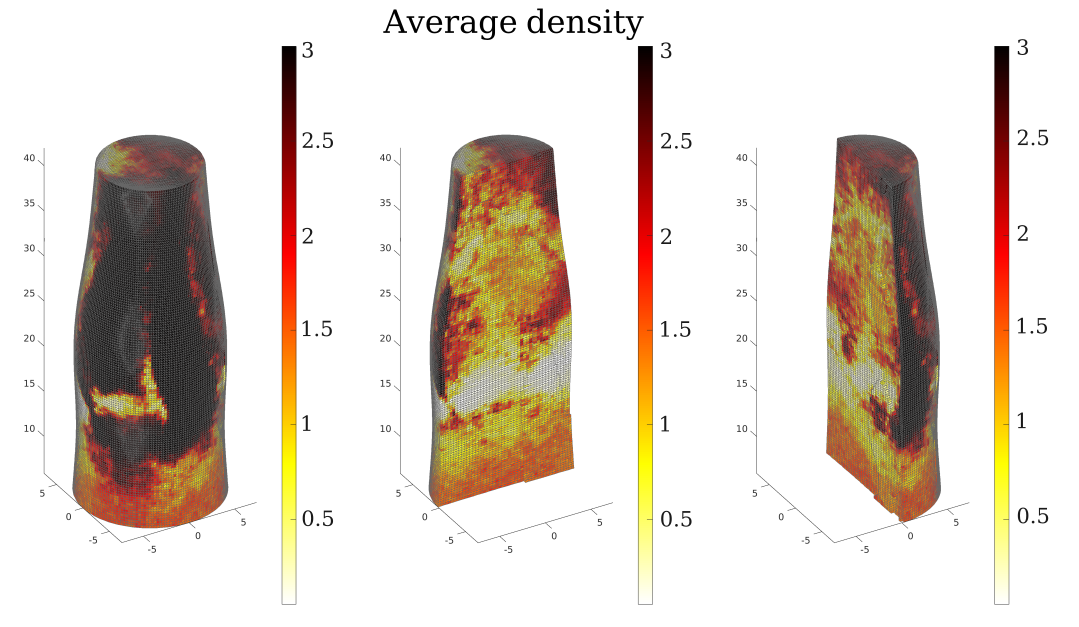}
	\includegraphics[width=\textwidth,trim=0.8cm 0cm 0cm 0cm,clip=true]{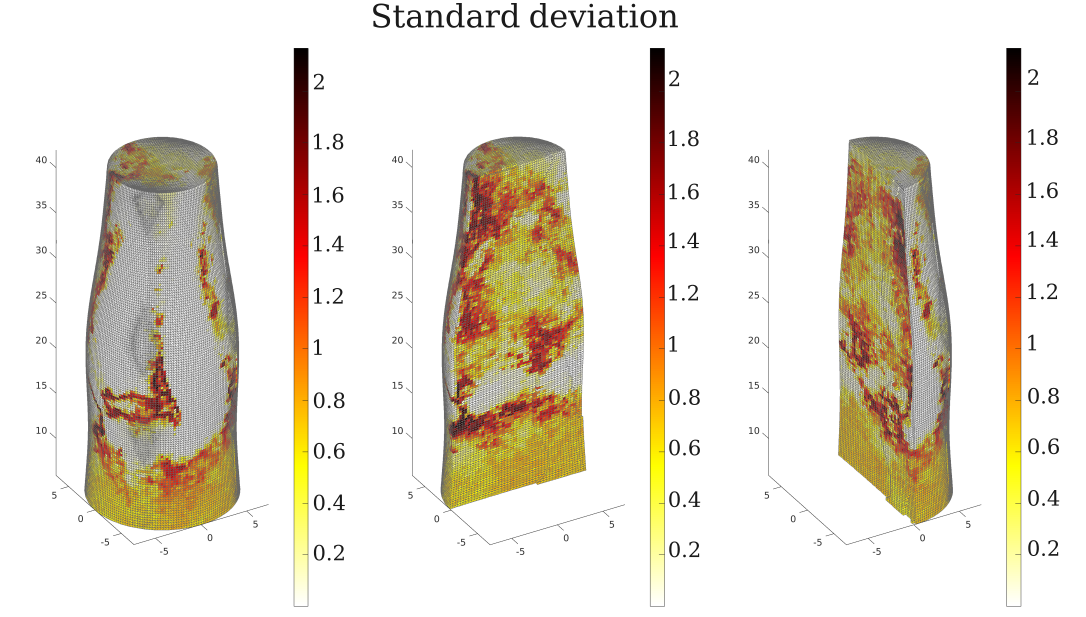}
	\caption{Three dimensional density profile ( top) and standard deviation ( bottom) of the blast furnace studied. The axes ( XYZ) are in meters and the density in \si{g.cm^{-3}}.\\
 - \textit{at the top} : distribution of the \textbf{average density} in the blast furnace, \\
	- \textit{at the bottom} : distribution of the \textbf{density standard deviation } in the blast furnace.}
	\label{corsikadata}
	
\end{figure}

The 3 muon runs were performed at different periods. Therefore, the 3D reconstruction of the density distribution of the blast furnace gives us an average of what we can find and not a snapshot of the different zones and their thickness. In figure \ref{corsikadata}, the distribution of the mean density (at the top) and its standard deviation (at the bottom) are shown using real data. The theoretical muon flux was taken from a model built with CORSIKA \citep{heck1998corsika,cohu2022atmospheric}.

 \begin{itemize}
     \item The shell of the blast furnace is clearly visible. It is very dense: more than 3~\si{g/cm^3}. A \textit{travel length} from the detector in position 2 crosses only the shell, so we have directly the value of the density at this point (from the measured flux), without integrate it over the whole width of the blast furnace. 
    
     \item A brighter area is noticeable on the shell (bottom left of the average density figures). This is probably an area where the three positions provide different information on the integrated density. The value of the associated standard deviations is also high, maybe one of the detectors cannot see this area it brings incertainty in reconstruction.
     
     \item We distinguish, at a height of 15-20 m inside the BF, a sparse zone ($<$0.5 \si{g/cm^3}) that would contain mainly coke/coal. 
     
     \item From this last zone would leave a slightly denser zone in the shape of a \textit{chimney}. This phenomenon appears when a lot of coal is pulverized in the center and little agglomerate. This is the case in the blast furnace that we have studied.
     
     \item The cohesive zone is visible at 20-25 m height, with a density higher than 1.5~\si{g/cm^3}. It is close to the shell and not spread to the middle .
     
     \item The superposition of materials in what we call the "dry zone" is not visible. In this zone, all materials charged from the top of the furnace are in the solid state. The iron charge and the coke are descending and maintain a layered structure. There is a superposition of coke and sinter sublayers with a density of 0.7 \si{g/cm^3} and 2 \si{g/cm^3} respectively. 
 \end{itemize}
 
 The results of the 3D inversion obtained here are very satisfactory because we see density contrasts between zones and the cohesive zone could be clearly observed, with a higher density than the rest (except the shell). The algorithm and the inversion method have been successfully tested on the internal structure of the BF. We will test the robustness of the inversion in the next subsection. Operators of the blast furnace studied found the 3D reconstruction consistent with reality.

 \subsection{Analysis of various sensitivities \label{sensi_data}}
First, we studied the differences observed on our 3D reconstructions as a function of the muon flux model. We compared Tang et al.\citep{tang_muon_2006} and CORSIKA to calculate the muon flux loss. Tang is an analytical model commonly used and the CORSIKA flux allows to adapt to the environmental conditions and to the location of an experiment \citep{cohu2022atmospheric}. We analyzed the performances and uncertainties of the reconstruction engine: what are the differences caused by the randomness of the inversion. Then we looked at the consequences of the number of models registered in the MCMC algorithm and the randomness of the algorithm itself. We also tested what is the effect of a different density model in the MCMC algorithm input (different density at the cohesive zone). In all cases, the systematic comparison of theoretical models, the input density model, number of models registered etc. do not influence the final result. The algorithm gives stable reconstructions : density values oscillates +/- 0.4$\%$ between 2 models and the standard deviation value for a voxel is close to 0.2. The areas with the largest density differences (std $\sim$ 0.8) are located at the bottom of the blast furnace: where little data is collected which is obviously a source of noise.



\section{Conclusion}

We performed a muography experiment on a blast furnace of ArcelorMittal and obtained the first 3D image of it. We clearly distinguished the location of the cohesive zone, which turns out to be the key of the furnace productivity. The 3D reconstruction of the density distribution was a great success. We used an inversion program on our measured muon data using Markov chain Monte Carlo. This algorithm is stable (few variations between 2 identical models) and rather fast even with a large number of recorded models: consistent for operators to see changes several times a day. The results of the 3D inversion with the use of CORSIKA or Tang \citep{tang_muon_2006} models of theoretical flux show that the opacity estimates are strongly influenced by the theoretical flux model choice in the regions of zenith angle between 70 and 90° and especially for the areas of low opacity. The theoretical flux of CORSIKA is also dependent on atmospheric conditions and is adapted to the season when the acquisitions were made. The 3D images obtained are only an average of the density distribution but are quite realistic and validated by the operators of the blast furnaces studied. The density contrasts are obvious, especially for the shell and the cohesive zone which are clearly distinguishable. We could monitor the activity inside the BF. We have evaluated the effect of the atmospheric pressure variation on the measured muon flux and we are able to correct it for this impact. The measured and corrected flux seems to be sensitive to the coke rate variations in the cohesive zone (especially during high pressure period). All these elements related to the composition or the shape of the cohesive zone may allow the blast furnace operators to adapt their material loading according to the state of this zone. Further improvements to the method are under study. \\

\acknowledgments

This work was the subject of a CIFRE agreement between ArcelorMittal Maizières Research SA and IP2I (Lyon).






\begin{thebibliography}{99}
\bibitem{takamatsu}
Takamatsu, Nobuhiko, et al. \emph{Development of iron-making technology.},\emph{Nippon Steel Technical Report 101} (2012): 79-88.
\bibitem{vanini}
Vanini, S., et al. \emph{Muography of different structures using muon scattering and absorption algorithms.}, \emph{Philosophical Transactions of the Royal Society} A 377.2137 (2019): 20180051.

\bibitem{nagamine}
Nagamine, K., Tanaka, H. K., Nakamura, S. N., Ishida, K., Hashimoto, M., Shinotake, A., ... ,Hatanaka, A. .  \emph{Probing the inner structure of blast furnaces by cosmic-ray muon radiography}.  \emph{Proceedings of the Japan Academy}, Series B, 81(7), (2005) 257-260.
\bibitem{hu}
Xianfeng Hu, Lena Sundqvist Ökvist, Elin Åström, Fredrik Forsberg, Paolo Checchia, Germano Bonomi, Irene Calliari, Piero Calvini, Antonietta Donzella, Eros Faraci, et al. \emph{Exploring the capability of muon scattering tomography for imaging the components in the blast furnace.} \emph{ISIJ International}, pages ISIJINT–2017, (2017).
\bibitem{rosas}
Rosas‐Carbajal, Marina, et al. \emph{Three‐dimensional density structure of La Soufrière de Guadeloupe lava dome from simultaneous muon radiographies and gravity data.} \emph{Geophysical Research Letters} 44.13 (2017): 6743-6751.

\bibitem{Nagahara}
Nagahara, Shogo, et al. \emph{Three-dimensional density tomography determined from multi-directional muography of the Omuroyama scoria cone, Higashi–Izu monogenetic volcano field, Japan.} \emph{Bulletin of Volcanology} 84.10 (2022): 94.
\bibitem{lesparre_design_2012}
Lesparre, N and Marteau, J and Déclais, Y and Gibert, Dominique and Carlus, B and Nicollin, Florence and Kergosien, Bruno, \emph{Design and operation of a field telescope for cosmic ray geophysical tomography}, \emph{Geoscientific Instrumentation, Methods and Data Systems, Copernicus GmbH} {\bf 1} (2012) 11.
\bibitem{tarantola2005inverse}
Tarantola, Albert, \emph{Inverse problem theory and methods for model parameter estimation}, \emph{SIAM}(2005).


\bibitem{sambridge2002monte}
Sambridge, M and Mosegaard, K, \emph{{Monte Carlo} methods in geophysical inverse problems}, \emph{Reviews of Geophysics, Wiley Online Library} {\bf 40} (2002) 3--1.

\bibitem{mosegaard1995monte}
Mosegaard, Klaus and Tarantola, Albert, \emph{{Monte Carl}o sampling of solutions to inverse problems}, \emph{Journal of Geophysical Research: Solid Earth, Wiley Online Library} {\bf 100} (1995) 12431--12447.

\bibitem{chevalier2014monte}
Chevalier, A and Legchenko, Anatoli and Girard, J-F and Descloitres, Marc,  \emph{{Monte Carlo} inversion of {3-D} magnetic resonance measurements}, \emph{Geophysical Journal International, Oxford University Press} {\bf 198} (2014), 216--228.


Geoscientific Instrumentation, Methods and Data Systems,
\bibitem{avgitas}
T Avgitas, S Elles, C Goy, Y Karyotakis, and J Marteau. \emph{Mugraphy applied to archaelogy}. \emph{arXiv preprint arXiv} :2203.00946, (2022).


\bibitem{acernese2022virgo}
Acernese, F and Agathos, M and Ain, A and Albanesi, S and Allocca, A and Amato, A and Andrade, T and Andres, N and Andr{\'e}s-Carcasona, M and Andri{\'c}, T and others, \emph{The Virgo O3 run and the impact of the environment}, \emph{arXiv preprint arXiv:2203.04014} (2022) .

\bibitem{marteau_muons_2012}
Marteau, J. and Gibert, D. and Lesparre, N. and Nicollin, F. and Noli, P. and Giacoppo, F., \emph{Muons tomography applied to geosciences and volcanology}, \emph{Nuclear Instruments and Methods in Physics Research Section A: Accelerators, Spectrometers, Detectors and Associated Equipment, Science Direct}(2012).



\bibitem{jourde}
Jourde, Kevin, Dominique Gibert, and Jacques Marteau. \emph{Joint inversion of muon tomography and gravimetry-a resolving kernel approach.} \emph{arXiv preprint arXiv:1411.5146} (2014).


\bibitem{tramontini2019middle}
Tramontini, Matias and Rosas-Carbajal, Marina and Nussbaum, Christophe and Gibert, Dominique and Marteau, Jacques, \emph{Middle-atmosphere dynamics observed with a portable muon detector}, \emph{Wiley Online Library} {\bf 6} (2019) 1865--1876.
\bibitem{adamson2014observation}
Adamson, P and Anghel, I and Aurisano, A and Barr, G and Bishai, M and Blake, A and Bock, GJ and Bogert, D and Cao, SV and Castromonte, CM and others, \emph{Observation of muon intensity variations by season with the MINOS near detector}, \emph{APS} (2014) 012010.



\bibitem{jourde_monitoring_2016}
Jourde, K and Gibert, D and Marteau, J and de Bremond d'Ars, J and Gardien, S and Girerd, C and Ianigro, JC, \emph{Monitoring temporal opacity fluctuations of large structures with muon radiography: a calibration experiment using a water tower}, \emph{Scientific Reports} {\bf 6} (2016).

\bibitem{heck1998corsika}
Heck, Dieter and Schatz, G and Knapp, J and Thouw, T and Capdevielle, JN, \emph{{CORSIKA: a Monte Carlo} code to simulate extensive air showers} (1998) .





\bibitem{cohu2022atmospheric}
Cohu, Am{\'e}lie and Tramontini, Matias and Chevalier, Antoine and Ianigro, Jean-Christophe and Marteau, Jacques, \emph{Atmospheric and Geodesic Controls of Muon Rates: A Numerical Study for Muography Applications}, \emph{Instruments, MDPI} {\bf  6} (2022) 24.

\bibitem{tang_muon_2006}
Tang, Alfred and Horton-Smith, Glenn and Kudryavtsev, Vitaly A. and Tonazzo, Alessandra, \emph{Muon simulations for {Super}-{Kamiokande}, {KamLAND}, and {CHOOZ}}, \emph{Physical Review D} {\bf 74} (2006) .









\end{thebibliography}
\end{document}